\documentclass[journal]{IEEEtran}
\usepackage{graphicx}
\graphicspath{ {./} }

\usepackage{array}
\usepackage{caption}
\usepackage{float}
\usepackage[caption = false]{subfig}

\usepackage[table]{xcolor} 

\usepackage{multirow}
\usepackage{amsmath}

\usepackage{tikz}
\usetikzlibrary{arrows.meta, positioning, shapes.multipart}

\ifCLASSINFOpdf
\else
\fi

\begin{document}

%\title{Heart murmur detection, classification and analysis using PCG datasets from Physionet challenge 2022  }
%\title{A Novel Method for Heart Murmur Detection and Analysis from Phonocardiogram Recordings }
%\title{Heart Murmur Detection from Phonocardiogram Recordings via a Shallow 2D-CNN }
\title{Heart Murmur and Abnormal PCG Detection via Wavelet Scattering Transform \& a 1D-CNN }


\author{Ahmed~Patwa,~\IEEEmembership{Member,~IEEE,}
        Muhammad~Mahboob~Ur~Rahman,~\IEEEmembership{Senior Member,~IEEE,}
        and~Tareq~Y.~Al-Naffouri,~\IEEEmembership{Senior Member,~IEEE}% <-this % stops a space
\thanks{Authors are with the CEMSE division, King Abdullah University of Science and Technology (KAUST), Saudi Arabia. This work is supported in part by the KAUST smart health initiative (KSHI) seed grant fund, and KAUST baseline fund BAS/1/1665-01-01.}} % <-this % stops a space
\maketitle

\begin{abstract}

%Heart murmurs provide valuable information about mechanical activity of the heart, and thus, are clinically significant measure which aid in diagnosis of various heart valve diseases. 

This work does automatic and accurate heart murmur detection from phonocardiogram (PCG) recordings. Two public PCG datasets (CirCor Digiscope 2022 dataset and PCG 2016 dataset) from Physionet online database are utilized to train and test three custom neural networks (NN): a 1D convolutional neural network (CNN), a long short-term memory (LSTM) recurrent neural network (RNN), and a convolutional RNN (C-RNN). We first do pre-processing which includes the following key steps: denoising, segmentation, re-labeling of noise-only segments, data normalization, and time-frequency analysis of the PCG segments using wavelet scattering transform. We then conduct four experiments, first three (E1-E3) using PCG 2022 dataset, and fourth (E4) using PCG 2016 dataset. It turns out that our custom 1D-CNN outperforms other two NNs (LSTM-RNN and C-RNN). 
Further, our 1D-CNN model outperforms the related work in terms of accuracy, weighted accuracy, F1-score and AUROC, for experiment E3 (that utilizes the cleaned and re-labeled PCG 2022 dataset). As for experiment E1 (that utilizes the original PCG 2022 dataset), our model performs quite close to the related work in terms of weighted accuracy and F1-score.

%Specifically, our 1D-CNN model achieves an accuracy of 74.39\%, weighted accuracy of 78.06\%, F1-score of 62.3\%, and area under receive operating characteristic (AUROC) curve of 82.11\% for experiment E1; an accuracy of 90.09\%, F1-score of 90.09\%, and AUROC of 95.32\% for experiment E2; an accuracy of 78.7\%, weighted accuracy of 83.69\%, F1-score of 78.67\%, and AUROC of 91.82\% for experiment E3; an accuracy of 96.51\%, F1-score of 96.42\% and AUROC of 99.1\% for experiment E4.

%we also abnormal PCG detection. relatively shallow 1D-CNN that consists of 4 convolutional layers and 4 fully connected layers. Specifically, our 1D-CNN does murmur detection with an accuracy of 79\%, and an area under the receiver operating characteristic (AUROC) of 0.84. Furthermore, the experiments were repeated excluding the low-quality recordings (annotated in CirCor dataset as "Unknown"), resulting in an even higher performance, with a classification accuracy of 86\% and an AUROC that is 0.93. Last but not the least, the proposed DL model was also used to make a distinction between the normal and abnormal PCG signals from CinC-Physionet-2016 dataset. The results obtained were really good, with a classification accuracy of 96\% and an AUROC that is 0.99.

\end{abstract}

\begin{IEEEkeywords}
phonocardiogram, heart murmur, valvular heart disease, physionet, convolutional neural network, recurrent neural network, wavelet scattering.
\end{IEEEkeywords}

% \fi
\IEEEpeerreviewmaketitle

\section{Introduction}

Cardiac auscultation provides valuable insights into the mechanical activity of heart, and remains the gold standard for diagnosis of a wide range of cardiovascular diseases. Phonocardiogram (PCG)---the systematic recording of heart sounds by means of a digital stethoscope, is a clinically significant method to study the pathology of the four heart valves (i.e., the mitral valve, tricuspid valve, aortic valve, and pulmonary valve). This is because the PCG could record abnormal sounds, known as heart murmurs, made by heart valves as they open and close during cardiac cycle. Heart murmurs, when present, often indicate the presence of a heart valve disease, e.g., mitral regurgitation, mitral stenosis, congenital heart disease, septal defects, patent ductus arteriosus in newborns, defective cardiac valves, rheumatic heart disease, etc \cite{review1}. Thus, early murmur detection, classification, and grading analysis in an automated fashion to help diagnose a valvular heart disease at an early stage is the need of the hour \cite{review2}. 

PCG signal acquisition is typically done from following four chest locations where the valves can be best heard: Aortic area, Pulmonic area, Tricuspid area, and Mitral area \cite{2022challengepaper}. For a normal sinus cardiac cycle, the PCG signal captures two fundamental heart sounds, first is called S1 sound while second is called S2 sound. S1 signifies the start of isovolumetric ventricular contraction with the closing of the mitral and tricuspid valves amid rapid increase in pressure within the ventricles. S2, on the other hand, implies the start of diastole with the closing of the aortic and pulmonic valves. In addition to S1 and S2 sounds, the mechanical activity of the heart may occasionally give rise other sounds which include: third heart sound (S3), the fourth heart sound (S4), systolic ejection click (EC), mid-systolic click (MC), diastolic sound or opening snap (OS), and heart murmurs (due to turbulent flow of blood due to malfunctioning of heart valves) \cite{2016DS}.   

PCG signal analysis for abnormal heart beat detection and for murmur detection has traditionally been done using classical signal processing techniques \cite{classicaldsp}. More recently, there has been enormous interest in utilizing tools from machine learning \cite{mlpcg}, deep learning \cite{dlpcg}, and transfer learning \cite{tlpcg} in order to do heart sounds classification (see the review articles \cite{review1},\cite{review2}, and references therein for more in-depth discussion of the related work). 

Inline with the recent surge of interest in utilizing deep learning methods for PCG signal analysis, this work utilizes two public PCG datasets (PCG 2022 dataset and PCG 2016 dataset) from Physionet database \cite{physionetwebsite} to train and test three custom neural networks (NN), i.e., 1D-convolutional neural network (CNN), long short-term memory (LSTM)-recursive neural network (RNN), convolutional-RNN (C-RNN) for heart murmur and abnormal PCG detection. 

{\bf Contributions.}
The main contributions of this work are as follows:
\begin{itemize}
\item {\it Dealing with the noisy datasets:} The two PCG datasets are occasionally corrupted with noise (e.g., baby crying, background sounds, etc.). Thus, we pre-process the two datasets in order to split each PCG recording into smaller segments using a custom-designed GUI framework. This way, we systematically identify and remove the noise-only segments. We also perform ablation study to determine the optimal window size for segmentation. 
\item {\it Dealing with small-sized datasets:} We note that both PCG datasets are small in size; therefore, we utilize {\it Wavelet scattering transform}--a tool that efficiently handles small datasets-- for feature extraction and for the time-frequency analysis on our denoised, pre-processed, segmented data. 
\item {\it Dealing with the imbalanced datasets:} We minimize the class imbalance in both datasets by reducing the size of the murmur absent class (by downsampling the murmur absent class by weighted random sampling). The weighted random sampling method assigns a certain weight to each instance in the training data, allowing instances from the minority class to be sampled more than once, which in turn leads to increasing the total number of instances from the minority class due to repetition.  
\item {\it Design of four experiments:} We design four experiments to evaluate the performance of our NN models on following four datasets: 1) PCG 2022 dataset as is, 2) PCG 2022 dataset with unknown class removed, 3) PCG 2022 dataset with noise-only segments removed, and 4) PCG 2016 dataset as is. 
\item {\it Voting-based approach:} For the experiments E1-E3, we also evaluate a voting based approach where we group all the samples that belong to the same heart auscultation location of a given patient. We then inspect the classification result of each sample, and finally choose the label that has the maximum number of votes. 
\end{itemize}

%Specifically, our 1D-CNN model achieves an accuracy of 74.39\%, weighted accuracy of 78.06\%, F1-score of 62.3\%, and area under receive operating characteristic (AUROC) curve of 82.11\% for experiment E1; an accuracy of 90.09\%, F1-score of 90.09\%, and AUROC of 95.32\% for experiment E2; an accuracy of 78.7\%, weighted accuracy of 83.69\%, F1-score of 78.67\%, and AUROC of 91.82\% for experiment E3; an accuracy of 96.51\%, F1-score of 96.42\% and AUROC of 99.1\% for experiment E4. 

{\it Performance advantage over the related work:}
Among the three models we have implemented, the vanilla 1D-CNN model outperforms the other two NN models (i.e., LSTM-RNN and C-RNN). 
Further, the vanilla 1D-CNN model outperforms the related work in \cite{rank1,rank2,rank3} in terms of accuracy, weighted accuracy, F1-score and AUROC, for experiment E3 (that utilizes the cleaned and re-labeled PCG 2022 dataset). As for experiment E1 (that utilizes the original PCG 2022 dataset), our model performs quite close to \cite{rank1} in terms of weighted accuracy, and stays quite close to \cite{rank2} in terms of F1-score.

{\bf Outline.}
The rest of this paper is organized as follows. Section II describes selected related work on murmur detection and heart valve disease classification. Section III outlines essential details of the two public datasets used, pre-processing done, and time-frequency domain methods for feature extraction. Section IV presents three deep learning classifiers for murmur detection and abnormal PCG detection. Section V discusses performance results for all three classifiers for both datasets. Section VI concludes the paper.

\section{Related Work}

Broadly speaking, the relevant literature on PCG signal analysis has attempted to solve following major problems, to date: i) identification of fundamental heart sounds (S1 and S2), also known as lub-dub sounds \cite{s1s2}, ii) abnormal PCG detection (by identifying sounds other than the normal lub-dub sounds) \cite{abnormalheartsound}, iii) heart sound classification (normal sounds, murmurs, extra clicks, artifacts) \cite{heartsoundclassification}, iv) heart murmur detection \cite{murmurdetect2022dataset}, v) heart valve disease classification (by means of heart murmur classification) \cite{heartvalvedisease}, and vi) PCG signal denoising methods \cite{denoisingpcg}. Lately, there is a work which attempts to do automatic murmur grading analysis \cite{2022-grading} at the patient-level, and has the potential to do disease staging and progression analysis. 

In terms of public datasets, apart from PCG 2022 dataset \cite{2022challengepaper} and PCG 2016 dataset \cite{2016DS} that are available on Physionet database, there are a few other public datasets too. These include: EPHNOGRAM \cite{ephnogram}, PASCAL, Michigan heart sound and murmur database (MHSDB) provided by the University of Michigan health system, and cardiac auscultation of heart murmurs database provided by eGeneral Medical Inc. \cite{2016DS}. On a side note, there exists a fetal PCG dataset that provides 26 recordings of fetal PCG collected from expecting mothers with singleton pregnancy during their last trimester \cite{fetal-pcg}. 

Since this work focuses mainly on automatic heart murmur and abnormal PCG detection, selected work on these two problems is discussed as follows. {\cite{rank1},\cite{rank2},\cite{rank3} utilize the PCG 2022 dataset, and implement a light-weight CNN, a parallel hidden semi-Markov model, and a heirarchical multi-scale CNN, respectively, for murmur detection and clinical outcome prediction.}
\cite{sig_proccessing-2016} utilizes PCG 2016 dataset, applies multiple time-frequency analysis methods, e.g., discrete wavelet transform, mel-frequency cepstral coefficients etc., along with a feedforward neural network in order to do abnormal PCG detection, achieving 97.1\% accuracy.
\cite{unsegmented-2016Dataset} computes short time Fourier transform of the PCG signals without doing any segmentation, implements a custom CNN, and achieves an accuracy of 95.4\%, and 96.8\%, on PCG 2016 dataset and PASCAL dataset, respectively. 
\cite{Samit:IEEETIM:2022} utilizes two PCG datasets to evaluate their custom CNN model, as well as two pre-trained models (i.e., VGGNet-16, ResNet-50) in order to do heart valve disease (HVD) classification. They claim to achieve an overall accuracy of 99\% to classify the following HVD: aortic stenosis, mitral stenosis, aortic regurgitation, mitral regurgitation, and mitral valve prolapse.
\cite{cnn-classification} utilized a private dataset (with data collected at PKU Muhammadiyah Yogyakarta Hospital, Indonesia), and implemented a custom CNN model in order to HVD classification for the following diseases: angina pectoris, congestive heart failure, and hypertensive heart disease. They reported an accuracy of 85\% for their HVD classification problem. 
\cite{pediatric-heart-disease} aimed at intelligent diagnosis of murmurs in pediatric patients with congenital heart disease. They applied segmentation method to extract the first and second heart sounds from the PCG signals, and used them as input to their feedforward neural network classifier, resulting in HVD detection accuracy of 93\%. 
\cite{2015-multimodel-features} utilizes the PASCAL dataset and aims to classify PCG signals into three classes (normal, systolic murmur, diastolic murmur). To this end, they extract various time-domain features, frequency-domain features, and statistical features, and pass them to three classifiers: k-NN, fuzzy k-NN and a feedforward neural network. They report a classification accuracy of 99.6\%, that is achieved by the fuzzy k-NN. 
\cite{2016-sec-place} utilizes the PCG 2016 dataset, extracts time-domain, frequency-domain, and time-frequency domain features from the PCG signals, and implements an ensemble of 20 feedforward neural networks in order to do abnormal PCG detection, achieving an overall accuracy of 91.5\%.
\cite{2016-third-place} utilizes the PCG 2016 dataset, extracts time-domain, frequency-domain and time-frequency domain features, does feature selection and implements a two-layer feedforward neural network for abnormal PCG detection, resulting in an accuracy of 85.2\%. Finally, \cite{MD-Via-DS} trains a custom neural network using a dataset that consisted of 34 hours of heart sound recordings. Their murmur detection algorithm results in a sensitivity of 76.3\%, specificity of 91.4\%. Furthermore, they report an increase in sensitivity to 90\% when they omit soft murmurs with grade 1.

\section{Datasets \& Data Pre-processing }

As mentioned earlier, this work utilizes two public PCG datasets: 1) CirCor Digiscope PCG dataset from PhysioNet (called PCG 2022 dataset, in this work) \cite{2022challengedescription}, 2) PCG dataset from PhysioNet (called PCG 2016 dataset, in this work). Below, we describe the key details of the two datasets, followed by the key pre-processing steps performed on each dataset.

\subsection{Datasets}
{\it 1) CirCor Digiscope PCG dataset (PCG 2022 dataset):}

This CirCor Digiscope PCG dataset from Physionet consists of PCG signals collected from pediatric subjects during the two mass screening campaigns in Brazil. The dataset consists of 3,163 audio files, collected from 942 patients, sampled at 4,000 Hz. The length of these recordings vary between 5 seconds to 65 seconds. There are a total of 2,391 recordings for the murmur absent class, 616 recordings for the murmur present class, and 156 recordings for the unknown class. Thus, the dataset is highly imbalanced given the number of examples for the murmur present and murmur absent class.

All the PCG recordings have been taken from one or more of 4 possible locations on the chest, namely: AV, MV, PV, TV (see Fig. 1). Additionally, some samples have been taken at another location, labeled as Phc. Moreover, some subjects had multiple recordings for the same chest location. Murmur labels are given both patient-wise, and chest-location-wise. This implies that we could get to know whether a patient has murmur or not, as well as if the dataset annotator detected murmur in each recording separately. 

{\bf Remark:} Since each PCG recording is obtained from one unique chest location (out of four chest locations for auscultation), this implies we do murmur detection “chest-location-wise”, and not “patient-wise”. That is, we deal with each PCG recording on its own, regardless of the patient it belongs to. 

%The recordings itself in each split are dealt with separately, not like if they belong to the same patient. 

%\footnote{Only the released training data was utilized in this work. That is, all description about the CirCor Digiscope 2022 dataset in this paper is describing the training set. } 

\begin{figure}[h]
    \centering
    \includegraphics[width=0.25\textwidth]{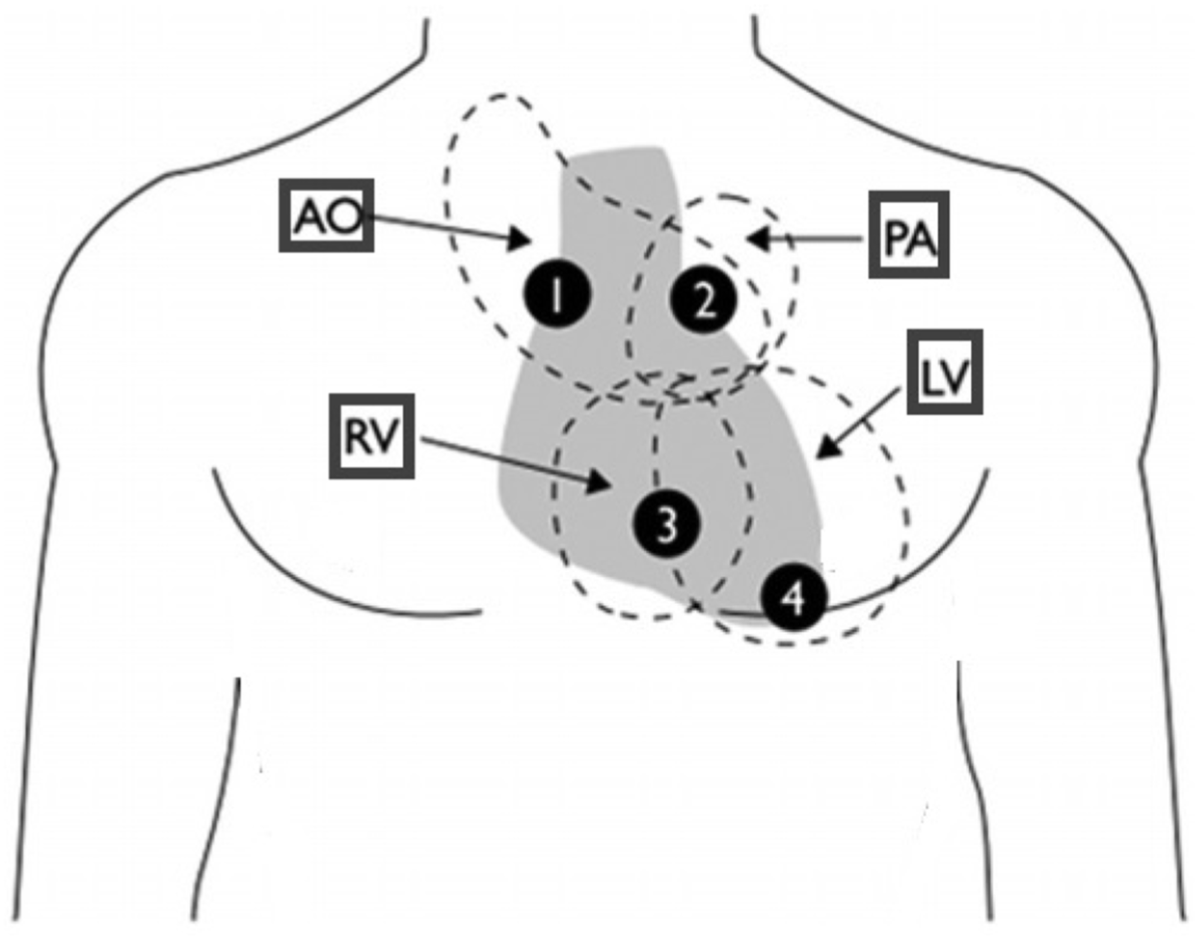}
    \caption{Cardiac auscultation locations: (PCG 2022 dataset) \cite{2022challengepaper}. }
    \label{fig:mesh1}
\end{figure}

%The length of each audio signal reaches more than 60 seconds in some cases. 
In addition to the audio PCG data, following meta-data is also available: age, weight, and gender of the pediatric subject. Furthermore, additional information about heart murmurs such as the location of a murmur, strength of a murmur is also given. Having said that, this work utilizes the CirCor Digiscope 2022 dataset for murmur detection, i.e., we solve a classification problem with following three classes: 1) murmur present, 2) murmur absent, and 3) unknown (noisy, unknown to the annotator).

{\it 2) PCG 2016 dataset:}

The PCG 2016 dataset from PhysioNet consists of a total of 3,240 heart sound recordings, sampled at 2,000 Hz. The length of these recordings vary between 5 seconds to 120 seconds. All the data was recorded from a single precordial location on the chest. As for the labels, each recording has been annotated as normal (taken from a healthy subject) or abnormal (taken from subjects with confirmed cardiac diagnosis). Furthermore, each recording has been labeled as either high-quality or low quality. Moreover, some of the data belonging to abnormal class has been further annotated with the exact diagnosis that the subject suffers from. There are a total of 2,575 recordings for the normal class, while there are a total of 665 recordings for the abnormal class. Thus, the data is highly imbalanced given the number of examples for the normal and abnormal class, and this effect is more pronounced when it comes to disease labels (within the abnormal class). 

We utilize PCG 2016 dataset to solve the classification problem with two classes: 1) PCG normal, 2) PCG abnormal. One important difference between this dataset and the CirCor Digiscope dataset is that this dataset assigns normal labels even to the low-quality recordings. Thus, these recordings could still be used for the classification problem (PCG normal vs. PCG abnormal), without adding a third class for low-quality recordings.  

{Table \ref{table:key-stats} summarizes the key statistics of the two datasets.}

\begin{table}[h]
    \caption{Key statistics of the two datasets}
    \centering
    \begin{tabular}{ |  >{\centering\arraybackslash}m{2.5cm} | >{\centering\arraybackslash}m{2.5cm} | >{\centering\arraybackslash}m{2.5cm }| }
        \hline
         & PCG 2022 dataset & PCG 2016 dataset\\
        \hline
        Total recordings     &   3,163           &   3,240              \\
        \hline
        Recordings per class   & Absent: 2,391, Present: 616,  Unknown: 156  & Normal: 2,575, Abnormal: 665   \\
        \hline
        Total subjects     &   942           &   -                     \\
        \hline
        Sampling rate          &  4,000 Hz          &  2,000 Hz         \\
        \hline
        Recordings lengths   &  (5-65) seconds & (5-120) seconds  \\
        \hline
        Auscultation locations   & AV, TV, MV, PV, Phc & Single chest location \\
        \hline
    \end{tabular}
    \label{table:key-stats}
\end{table}

\subsection{Data Pre-Processing}
{\it Denoising:}
Both datasets have noise overlapping with heartbeats. Some prominent examples of noise in the two datasets include: noise due to stethoscope movement, baby crying, intense breathing, and people talking. Since the spectrums of voice and PCG overlap, complete elimination of noise is not possible. However, out-of-band (OOB) noise rejection is still possible. Inline with the state of the art \cite{rank3}, a {butterworth} low-pass filter {of order 5} and a cut-off frequency of 500 Hz is applied to each PCG signal to remove the OOB noise.

{\it Segmentation:}
Since recordings are not of the same length, each recording has been divided into $N$-second long segments\footnote{The optimal segment length turns out to be $N=4$ seconds, as will be discussed in Section V in more detail.}, to increase the number of training examples for the deep learning model.
Thus, for PCG 2016 dataset, we end up with 12,827 PCG segments for the normal class and {3,920} PCG segments for the abnormal class. 
To deal with the problem of highly imbalanced dataset, a balanced subset is constructed by including all the PCG segments from the abnormal class, while randomly selecting PCG segments from the normal class. This way, {the two classes together} contain a total of 7,844 audio files (segments).  
The same process is repeated for the CirCor Digiscope 2022 dataset, where a total of 16,522 segments were extracted. We then utilized {6,822} files to construct as a more-balanced dataset as follows: {3,441} files from murmur absent class, 2,593 files from murmur present class, and {788} samples from the unknown class.

{\it Re-labeling of noise-only segments: }
The splitting of a PCG time-series into multiple smaller segments results in a new problem (i.e., how to assign labels to each segment). Note that a label (murmur present, murmur absent, unknown) was assigned to each recording as a whole by the annotator in the original dataset. Furthermore, when dividing a PCG signal into smaller segments, a situation arises where some segments are bad segments (i.e., they contain only noise, with no heartbeat at all!). If left untreated, feeding such noise-only segments to the deep learning model could have a negative impact on its performance. That is, the model will consider one such segment as representing a murmur. But in reality, the segment contains pure noise, and thus, should ideally be labeled as an unknown sample. Thus, to deal with this problem, a systematic method for assessment and re-labeling of each segment of the CirCor 2022 dataset is needed. 

\begin{figure}[htp]
    \centering
    \includegraphics[width=0.48\textwidth]{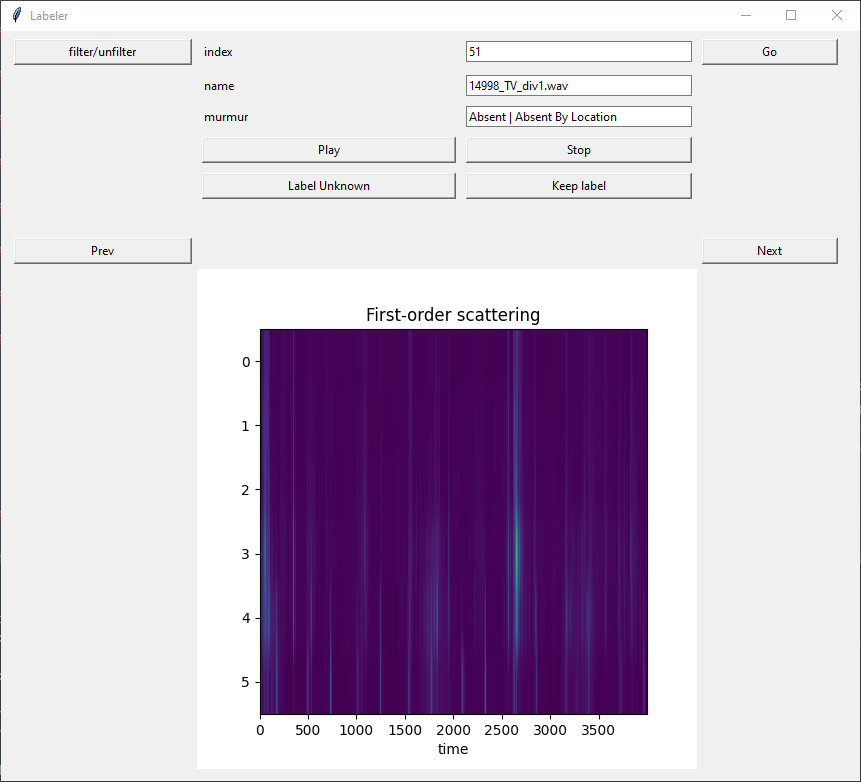}
    \caption{ Graphical user interface for guided re-labeling of dataset segments. }
    \label{fig:annotator}
\end{figure}

{\it Graphical user interface for segment assessment and potential re-labeling:}
We developed a simple graphical user interface (GUI) in Python, in order to assess the quality of the default label for each segment in order to re-label all the noise-only segments, in a semi-automated fashion (see Fig. \ref{fig:annotator}). Specifically, the GUI of Fig. \ref{fig:annotator} shows the 2D time-frequency domain representation of each segment, as well as plays the audio. This allows us to systematically determine the presence or absence of heart beat in each segment, for all PCG files. Note that the re-labeling process made sure not to alter the integrity of the dataset. Specifically, the only allowed change was to re-label (murmur present) to (unknown), or (murmur absent) to (unknown) indicating that a certain segment of a whole recording doesn't actually represent a heartbeat signal, but rather contains only noise. In other words, no (murmur present) was changed to (murmur absent) or the other way around. Similarly, no (unknown) was changed to (murmur present) or (murmur absent). This exercise helped us identify the bad segments, which in turn helped us improve the training accuracy of our proposed DL classifiers. The resulting new distribution of the three classes (murmur present, murmur absent, and unknown) is summarized in Table II. 
%The interface is shown in figure \ref{fig:annotator}.

\textbf{}

\begin{table}[h]
    \centering
    \caption{\footnotesize{CirCor Digiscope 2022 dataset: new distribution of the three classes after segment re-labeling}}
    \label{tab:relabeling}
    \begin{tabular}{>{\centering\arraybackslash}m{2.5cm}>{\centering\arraybackslash}m{1cm}>{\centering\arraybackslash}m{1cm}>{\centering\arraybackslash}m{1cm}}
        \hline
        
        &\multicolumn{3}{ c }{\textbf{Class}} \\ 
        \hline 
         & \textbf{Absent} & \textbf{Present} & \textbf{Unknown} \\
        \hline
        Before re-labeling & {3,441} & 2,593 & {788} \\
        After re-labeling & {2,363} & 2,438 & {1,976} \\
        \hline
    \end{tabular}
\end{table}

\begin{figure*}[htp]
    \centering
    \includegraphics[width=7in]{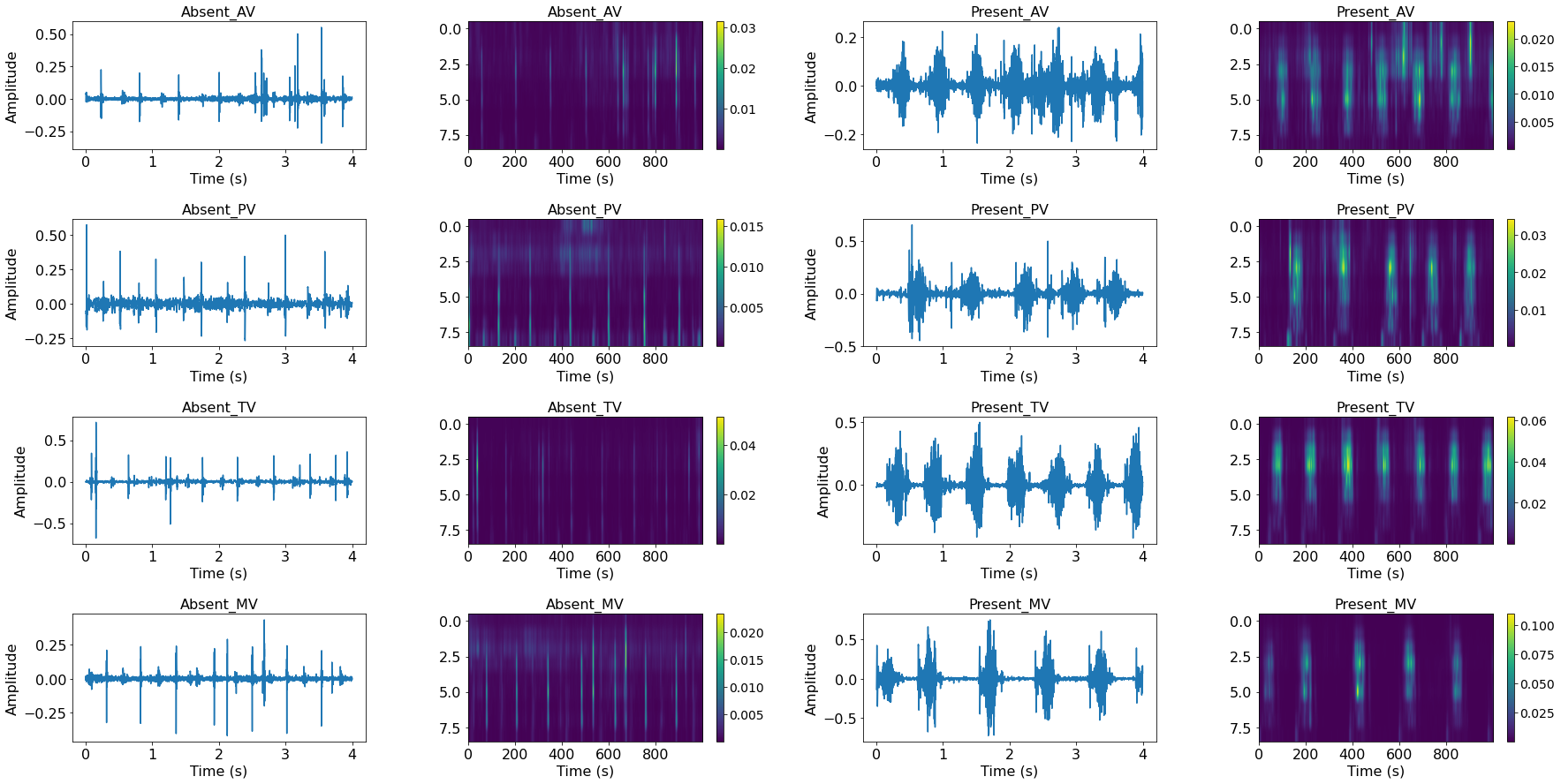}
    \caption{ PCG time-series and the corresponding WST for the two situations (murmur present, murmur absent) across the four heart valves, i.e., AV, PV, TV, MV (for CirCor Digiscope 2022 dataset). Note that WST is quite effective in differentiating between the two classes (murmur present, murmur absent). }
    \label{fig:murmur presence}
\end{figure*}

{\it Data normalization:}
For each data vector (segment), its mean $\mu$ and standard deviation $\sigma$ is computed. Then, i-th element $x_i$ of each data vector is normalized by means of following operation: $x_{i,n}=\frac{x_i-\mu}{\sigma}$. The normalized data vector has zero mean and unit variance.

\section{Time-Frequency Analysis for Feature Extraction}

In order to study the feasibility of the two classification problems at hand, and in order to help our neural networks do automatic feature extraction, we transform the 1D audio PCG signals into their equivalent 2D representation using three time-frequency domain methods: 1) short-time Fourier transform, 2) Mel-frequency Cepstral coefficients, and 3) Wavelet scattering transform. 

\subsection{Short-Time Fourier Transform}
The Short-Time Fourier Transform (STFT) analyzes the frequency content of localized sections of a signal as it changes over time. That is, STFT works by dividing a signal in time domain to shorter equally-sized segments, then computing Fourier transform on each segment. The STFT of a signal \( x(t) \) is defined as:
\begin{equation}
    \text{STFT}\{x(t)\}(t, \omega) = X(t, \omega) = \int_{-\infty}^{\infty} x(\tau) w(\tau - t) e^{-j \omega \tau} d\tau,
\end{equation}
where \( w(t) \) is the window function, \( \omega \) is the angular frequency, \( t \) is the time-shift parameter, and \( \tau \) is the time variable.

For the PCG signals analysis, we implement the discrete version of the STFT:
\begin{equation}
    X[m, \omega_k] = \sum_{n=-\infty}^{\infty} x[n] w[n - m] e^{-j \omega_k n},
\end{equation}
where \( x[n] \) is the discrete-time PCG signal, \( w[n] \) is the discrete window function, \( m \) is the time index, and \( \omega_k \) is the discrete angular frequency.

\subsection{Mel-Frequency Cepstral Coefficients}

The Mel-frequency Cepstral coefficients (MFCC)---a popular speech and audio processing method---analyzes the short-term power spectrum of a sound signal on a Mel scale of frequency. Specifically, to obtain the MFCC features, the power spectrum obtained from the fast Fourier transform is passed to the following blocks: Mel Filter bank, logarithmic scaling, and discrete Cosine transform (see Appendix A, for more details). 

\subsection{Wavelet scattering transform}

The wavelet scattering transform is a multi-resolution analysis technique that builds on the continuous wavelet transform, as follows. The continuous wavelet transform of a signal \( x(t) \) with a wavelet \( \psi(t) \) is given by:
\begin{equation}
    W_x(a, b) = \frac{1}{\sqrt{a}} \int_{-\infty}^{\infty} x(t) \psi^*\left(\frac{t - b}{a}\right) dt,
\end{equation}
where \( a \) is the scale parameter, \( b \) is the translation parameter, and \( \psi^* \) is the complex conjugate of the wavelet function.

Next, the wavelet scattering transform involves applying wavelet transforms, followed by non-linear operations such as the modulus and averaging. In general, for a given signal \( x(t) \), the wavelet scattering transform can be expressed as a set of coefficients \( S x(t) \) derived iteratively by:
\begin{equation}
    S x = \{ S_0 x, S_1 x, S_2 x, \ldots \},
\end{equation}
where the zero-order scattering coefficient is defined as:
\begin{equation}
    S_0 x = x \star \phi_J,
\end{equation}
where \( \phi_J \) is a low-pass filter at scale \( J \). Next, the first-order scattering coefficient is defined as:
\begin{equation}
    S_1 x(t) = |x \star \psi_{\lambda_1}| \star \phi_J,
\end{equation}
where \( \psi_{\lambda_1} \) is the wavelet at scale \( \lambda_1 \). Similarly, the second-order scattering coefficient is:
\begin{equation}
    S_2 x(t) = ||x \star \psi_{\lambda_1}| \star \psi_{\lambda_2}| \star \phi_J,
\end{equation}
where \( \psi_{\lambda_2} \) is the wavelet at scale \( \lambda_2 \). In general, the $m$-th-order scattering coefficient is defined as:
\begin{equation}
    S_m x = \left| \left| \cdots \left| x \star \psi_{\lambda_1} \right| \star \psi_{\lambda_2} \cdots \right| \star \psi_{\lambda_m} \right| \star \phi_J.
\end{equation}

From the AI perspective, wavelet scattering transform is a technique that applies mathematical operations similar to those applied in convolutional layers of a neural network. Specifically, it utilizes wavelets as fixed convolutional filter, followed by a non-linearity (modulus operator), followed by averaging in order to find a low-variance representation of the data. This way, wavelet scattering transform produces representations that are invariant to translations and stable to deformations such as scaling and elastic distortions. Wavelet scattering transform is known to be effective for feature extraction, even for very small datasets. 

Fig. \ref{fig:murmur presence} demonstrates that wavelet scattering transform is indeed effective in discriminating between the two classes, for the CirCor Digiscope 2022 dataset. That is, Fig. \ref{fig:murmur presence} provides us a way to visually differentiate between the two classes (murmur present, murmur absent) at each of the four heart valves (by plotting the audio PCG time-series as well as the wavelet scattering function for all possible scenarios). 

%Typically, third party libraries in programming languages provide functions to compute STFT and MFCC, like the Pytorch torch.stft function, and torchaudio.transform.MFCC. 

%short-time Fourier transform. This technique simply applies the transform over short windows of the whole signal, which results when concatenated in a 2D time-frequency representation of that signal. n-fft and hop-length. For an acoustic signal of 4 seconds and 4000 sampling rate, the best representation that gives balanced time-frequency resolution resulted of using n-fft of 64 samples and hop-length of 16 samples. Mel-spctrograms serve the same purpose, with one difference, which is that frequencies are converted to a mel-scale, which mimics the way a human ear perceives acoustic signals. 

%\section{PCG to ECG classification}

%\subsection{w-GAN method}
%Previous works [] has introduced methods for PPG to ECG translation. One of these works [] claims that their model is capable of translating any correlated pair of signals. We put their claim to the test by training the model to translate PCG signals to ECG signals.

\section{Heart Murmur \& Abnormal PCG Detection}

\subsection{Neural Network Architectures}
\begin{figure*}[h]
    \centering
    \includegraphics[width=1\textwidth]{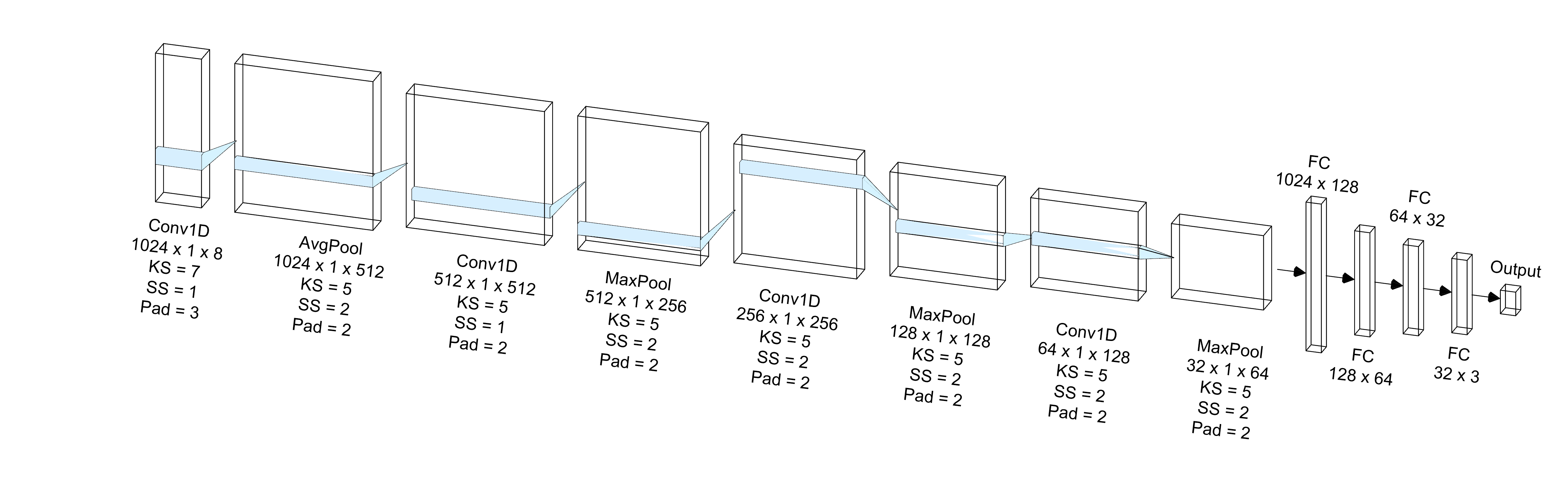}
    \caption{1D-CNN Model architecture: KS (kernel size), SS(stride size), Pad (padding). }
    \label{fig:CNN1D-Diagram}
\end{figure*}

For heart murmur detection and abnormal PCG detection, we implemented and tested the following neural network (NN) architectures: convolutional neural network (CNN)--both 1D and 2D, recurrent neural network (RNN)\footnote{The motivation behind using the RNN (LSTM and GRU) is to exploit its capability to learn temporal dependencies between the samples of a time-series (audio PCG signals, in this work).}, and a convolutional RNN (C-RNN). Eventually, it was the 1D-CNN model that resulted in maximum classification accuracy. Therefore, Fig. \ref{fig:CNN1D-Diagram} provides a detailed block diagram of the custom 1D-CNN model, with important parameters for each layer specified. One could see that the best-performer 1D-CNN model is relatively shallow, with 4 convolution layers (each followed by a pooling layer, followed by a batch normalization layer), a flattening layer, and 4 dense (fully connected) layers. Each (convolutional and dense) layer utilized RELU activation function, except the last layer which utilized softmax activation function. 

\begin{figure*}[h]
    \centering
    \includegraphics[width=1\textwidth]{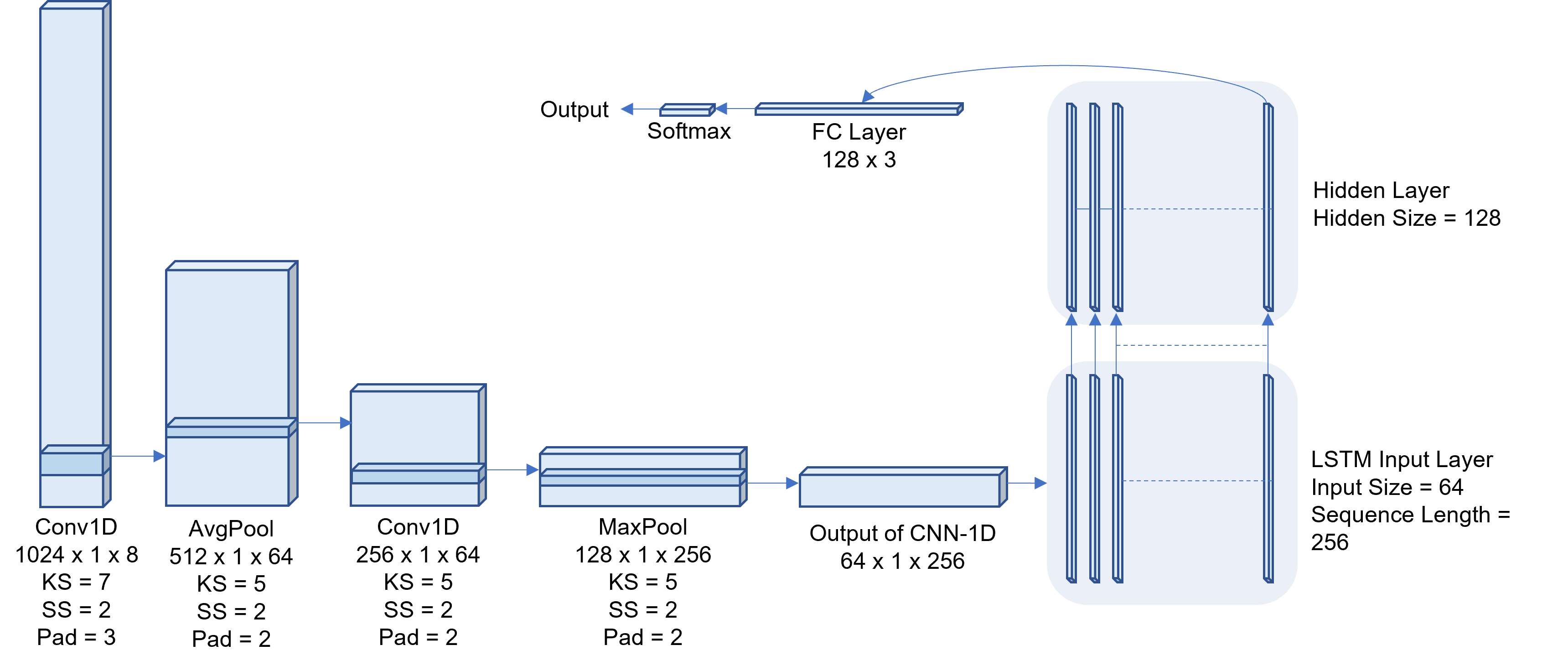}
    \caption{CRNN Model architecture: KS (kernel size), SS (stride size), Pad (padding). }
    \label{fig:CRNN-Diagram}
\end{figure*}

\begin{figure}[h]
    \centering
    \includegraphics[width=0.5\textwidth]{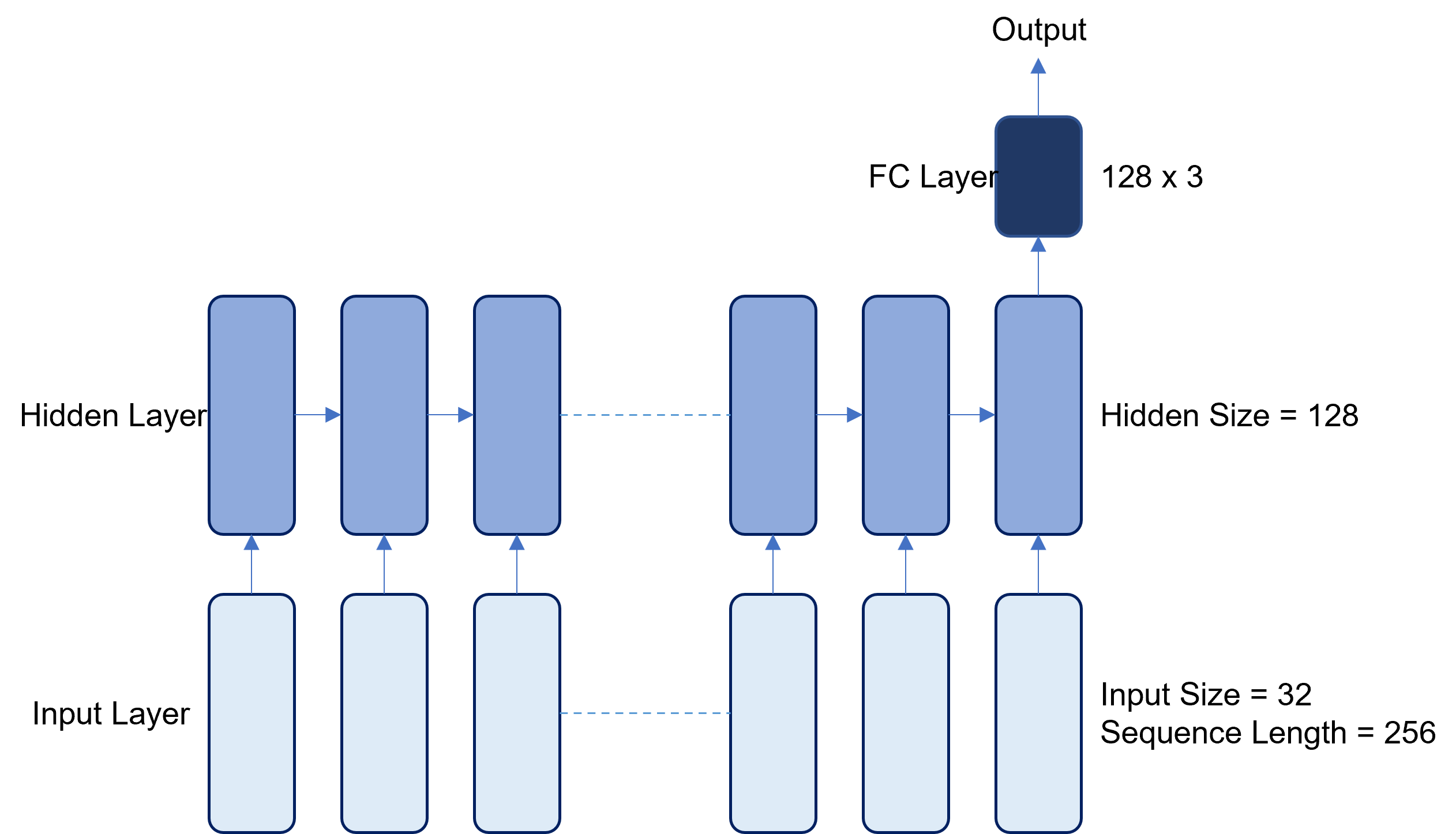}
    \caption{LSTM-RNN Model Architecture.}
    \label{fig:LSTM-Diagram}
\end{figure}

\begin{figure*}[htp]
    \centering
    \includegraphics[width=7in]{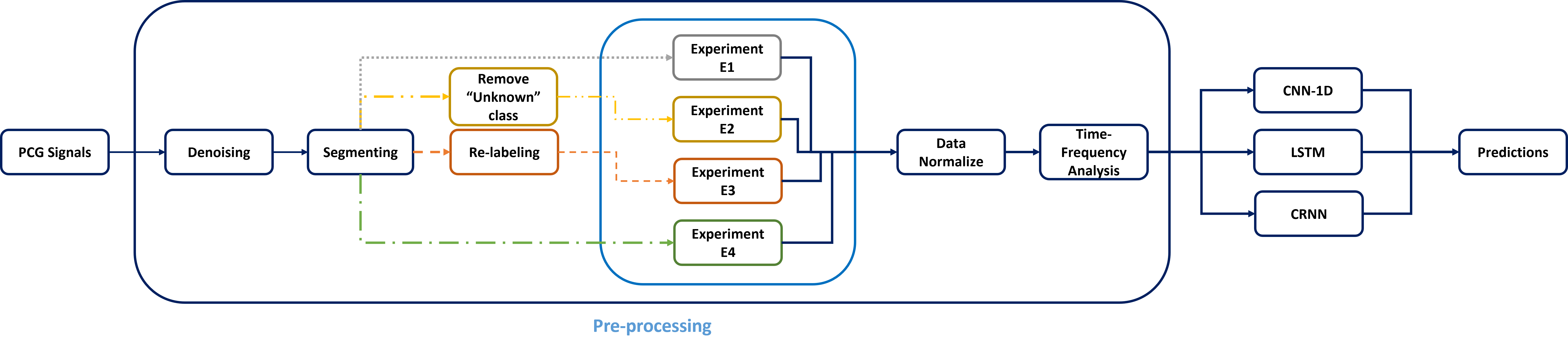}
    \caption{ Proposed method for heart murmur and abnormal PCG detection. }
    \label{fig:methodology}
\end{figure*}

For the sake of completeness, Figs. \ref{fig:CRNN-Diagram}, \ref{fig:LSTM-Diagram} present the block diagrams of the two other neural network models implemented in this work: a C-RNN model, and a single-layer LSTM-RNN model (with tanh activation function). Finally, Fig. \ref{fig:methodology} provides the overall pictorial summary of our approach for heart murmur and abnormal PCG detection.

Note that we re-train our 1D-CNN model from scratch in order to do abnormal PCG detection on the PCG 2016 dataset later\footnote{We investigated transfer learning technique by reusing the 1D-CNN model trained on PCG 2022 dataset by testing it on PCG 2016 dataset, but without much success. This points to the fundamental different nature (distribution) of the two datasets.}.

%For the purpose of murmur detection, demographic features were neglected, and the only information used is the three classes of murmur presence, in addition to the acoustic information extracted from the recordings itself.

\subsection{Training and Testing of NNs \& Hyper-parameters}

We implemented the custom 1D-CNN, the LSTM-RNN, and C-RNN models in Python using the PyTorch framework (in Microsoft visual studio environment), on a desktop PC with I5 7600 intel CPU and 1060ti Nvidia graphics card. {The total available data for each of the two datasets was divided into training data, validation data, and test data using a 70-15-15 split, while making sure that the test data is not seen by the models during the training stage.} For backpropagation purpose, the cross-entropy loss function was used. {To deal with class imbalance problem for both datasets, the "weighted random sampling" function provided in the Pytorch framework was used (in addition to downsampling, as discussed before). The weighted random sampling method assigns a certain weight to each instance in the training data, allowing instances from the minority class to be sampled more than once, which in turn leads to increasing the total number of instances from the minority class due to repetition. Specifically, the weight is calculated as follows: $W(class) = 1/n(class)$, where $W(class)$ is the weight for a certain class and $n(class)$ is number of instances in that class.}

Table \ref{table:HyperParameters} summarizes the important hyper-parameters for the 1D-CNN model and the LSTM-RNN model when trained on both datasets.

\begin{table}[h]
    \caption{\footnotesize{Hyper-parameters of the 1D-CNN and LSTM-RNN models when trained on both datasets. Q and J are wavelet scattering related parameters as described in the Python library Kymatio \cite{Kymatio}. For MFCC, nFFT is the size of window over which FFT will be applied, hop length is how much the window moves for each subsequent FFT, nMFCC is number of Cepstral coefficients to keep. (*A dropout of {0.25 drops 25}\% of the output of each dense layer, except last one.) } }
    \centering
    \begin{tabular}{ | >{\centering\arraybackslash}p{1.5cm} | >{\centering\arraybackslash}p{2cm} >{\centering\arraybackslash}p{2cm} >{\centering\arraybackslash}p{1.5cm}| }
        \hline
        Dataset/Model     & PCG 2022/ 1D-CNN & PCG 2022/ LSTM-RNN & PCG 2016/ 1D-CNN   \\ \hline
        Batch Size      & {126}        &    {126}                   &    126        \\ \hline
        Learning rate   & 3e-5       &    1e-3                  &    1e-4       \\ \hline
        Dropout         & {0.25*}     &      None                & {0.25*}          \\ \hline
        Optimizer       & Adam     &      Adam                & Adam          \\ \hline
        Backprop. algorithm         & SGD     &      SGD                & SGD          \\ \hline
        Initializer         & Xaviar     &      Xaviar                & Xaviar          \\ \hline
        Q               & 2          &          2               & 2             \\ \hline
        J               & 4          &          4               & 4             \\ \hline
        nFFT            & 128        &          -               & 128 \\ \hline
        hop length      & 16         &          -               & 16 \\ \hline
        nMFCC           & 10         &          -               & 10 \\ \hline
    \end{tabular}
    \label{table:HyperParameters}
\end{table}

We conducted a total of four experiments by training the custom 1D-CNN, the LSTM-RNN, and C-RNN on the two datasets, in order to do: 
\begin{itemize}
    \item E1) murmur detection using the original PCG 2022 dataset with 3 classes (murmur present, murmur absent, unknown).
    \item E2) murmur detection using a subset of the PCG 2022 dataset with 2 classes (murmur present, murmur absent) with unknown class excluded. 
    \item E3) murmur detection with 3 classes using the cleaned PCG 2022 dataset after re-labeling of noise-only segments, as discussed in previous section. 
    \item E4) abnormal PCG detection using the PCG 2016 dataset. 

\end{itemize}

{Table \ref{table:class-dist} shows the distribution of each class during the {\it test} split for each of the four experiments E1-E4.}

\begin{table}[h]
    \centering
    \caption{Class distribution in Test split for each experiment}
    \begin{tabular}{|c|c|c|c|}
        \hline
        \textbf{Experiment/Class} & \textbf{Absent} & \textbf{Unknown} & \textbf{Present} \\
        \hline
        E1 & 50.5\% & 11.5\% & 38\% \\
        \hline
        E2 & 49\% & - & 51\% \\
        \hline
        E3 & 24\% & 39\% & 37\% \\
        \hline
          & \textbf{Normal} & \textbf{Abnormal} & - \\
        \hline
        E4 &  58.5\% &  41.5\% & - \\
        \hline
        \end{tabular}
    \label{table:class-dist}
\end{table}

%{\it Loss and accuracy vs. epochs:} Fig. \ref{fig:murmur3-loss-acc} shows that the loss as well as the accuracy settles between 40 to 60 epochs during both the training and the validation phase, for experiment E1. Fig. \ref{fig:murmur2-loss-acc} shows that the loss and accuracy settle much quickly (in about 25 to 40 epochs), for experiment E2. This is due to the removal of the unknown class with noisy samples. The loss and accuracy plots for experiments E3 and E4 show a similar trend, and thus, are omitted for the sake of brevity.

%\begin{figure}[htp]
%    \centering
%    \subfloat[]{\includegraphics[width = 1.8in]{Loss-And-Accuracy Plots/Rev1/Murmur3Classes/Loss.png}} 
%    \subfloat[]{\includegraphics[width = 1.8in]{Loss-And-Accuracy Plots/Rev1/Murmur3Classes/Accuracy.png}}\\
%    \caption{ Loss and accuracy vs. epochs for experiment E1. }
%    \label{fig:murmur3-loss-acc}
%\end{figure}

%\begin{figure}[htp]
%    \centering
%    \subfloat[]{\includegraphics[width = 1.8in]{Loss-And-Accuracy Plots/Rev1/Murmur2Classes/Loss.png}} 
%    \subfloat[]{\includegraphics[width = 1.8in]{Loss-And-Accuracy Plots/Rev1/Murmur2Classes/Accuracy.png}}\\
%    \caption{ Loss and accuracy vs. epochs for experiment E2. }
%    \label{fig:murmur2-loss-acc}
%\end{figure}

\section{Experimental Results}

We first describe the performance metrics used to evaluate the 1D-CNN and other two classifiers, followed by a discussion of the key results for the three experiments E1, E2, E3 on murmur detection using the three variants of the PCG 2022 dataset, followed by a discussion of selected results for the fourth experiment E4 on abnormal PCG detection using PCG 2016 dataset. 

\subsection{Performance Metrics}

We use accuracy and F1-score as the main performance evaluation metrics\footnote{Since accuracy is most meaningful for a balanced dataset, we minimized the class imbalance by reducing the size of the murmur absent class in both datasets (by downsampling the murmur absent class by weighted random sampling).}. Additionally, weighted accuracy is also used as a performance metric, for the PCG 2022 dataset.

\subsubsection{Accuracy}
The accuracy $A$ of a classifier is the ratio of number of correct predictions $S_c$ by the model to the total number of samples $S_t$ in a dataset, i.e., $A = \frac{S_c}{S_t} \times 100 $. 

\subsubsection{Weighted Accuracy}
The weighted accuracy is a metric that gives more weight to the patients with murmur, and is defined as:

$A_w = \frac{5m_{pp} + 3m_{uu} + m_{aa}}{5(m_{pp} + m_{up} + m_{ap}) + 3(m_{pu} + m_{uu} + m_{au}) + (m_{pa} + m_{ua} + m_{aa})}$
where $m_{xx}$ is defined as in Table \ref{table:murmur_expert}.

\begin{table}[h]
    \centering
    \caption{Weighted accuracy confusion matrix}
    \label{table:murmur_expert}
    \begin{tabular}{|c|c|c|c|c|}
        \cline{3-5}
        \multicolumn{2}{c|@{}}{} & \multicolumn{3}{c|@{}}{\textbf{Murmur Classifier}} \\
        \cline{3-5}
        \multicolumn{2}{c|@{}}{} & \textbf{Present} & \textbf{Unknown} & \textbf{Absent} \\
        \hline
        \multirow{3}{*}{\textbf{Murmur Expert}} & \textbf{Present} & $m_{pp}$    & $m_{pu}$   & $m_{pa}$\\
        \cline{2-5}
        & \textbf{Unknown}                                         & $m_{up}$   & $m_{uu}$   & $m_{ua}$\\
        \cline{2-5}
        & \textbf{Absent}                                          & $m_{ap}$   &  $m_{au}$ & $m_{aa}$ \\
        \hline
    \end{tabular}
\end{table}

\subsubsection{F1-Score}
F1-score is a statistical measure that depends upon two factors, precision and recall. Precision is obtained by dividing total number of correctly classified elements, i.e. True Positives by total positively classified elements, i.e. True Positives (TP) + False Positives (FP). Thus, $P = \frac{TP}{TP + FP}$. Recall is obtained by dividing total number of positively classified samples by the total number of samples that should had been marked as positive, i.e., True Positives and True Positives + False Negatives (FN) respectively. Thus, $R = \frac{TP}{TP + FN} $.
With precision and recall in hand, F1-score is calculated as: $F1 = 2 * \frac{P*R}{P+R}$, where $P$ is precision and $R$ is recall.

% \begin{figure}[htp]
%     \centering
%     \subfloat[]{\includegraphics[width = 1.2in]{Confusion-Matrices/Rev1/cf_murmur_3classes.png}}
%     \subfloat[]{\includegraphics[width = 1.2in]{Confusion-Matrices/Rev1/cf_murmur_2classes.png}}
%     \subfloat[]{\includegraphics[width = 1.2in]{Confusion-Matrices/Rev1/cf_murmur_3classes_reannotated.png}}\\
%     % \includegraphics[width=9cm]{cf_murmur_2classes.png, cf_murmur_3classes.png}
%     \caption{ confusion matrices:\\(a) experiment E1: murmur detection with three classes,\\(b) experiment E2: murmur detection with two classes,\\(c) experiment E3: murmur detection with three classes (after re-labeling). }
%     \label{fig:cf-2022}
% \end{figure}

\begin{table}[h!]
\centering
\caption{Confusion matrices of the custom 1D-CNN model (for experiments E1, E2, and E3) }
\label{table:E1-E3}
\begin{tabular}{|c| c| c| c|c|} 
 \hline
 Experiment &
	Actual /Predicted&	Absent &	Unknown	& Present
\\

\hline
     &	Absent  &	\cellcolor{blue!25}82.56\% &                     10.27\% &	                   7.17\%  \\
E1 &	 Unknown &	                    74.58\% &	\cellcolor{blue!25}18.64\% &	                  6.78\%     \\
     &	Present &	                   9.77\%  &	                  3.60\% &	\cellcolor{blue!25}86.63\% \\
 
\hline
     &	Absent  &	\cellcolor{blue!25}92.03\% &                           - &	                   7.97\%  \\
E2 &	 Unknown &	                          - &	                        - &	                          - \\
     &	Present &	                   11.75\% &                           - &	\cellcolor{blue!25}88.25\% \\
     
\hline
     &	Absent  &	\cellcolor{blue!25}77.51\% &                      20.67\% &	                   1.82\% \\
E3 &	 Unknown &	                    19.75\% &	\cellcolor{blue!25}78.03\% &	                  2.23\% \\
     &	Present &	                   8.16\% &	                      11.32\% &	\cellcolor{blue!25}80.53\% \\

 \hline
\end{tabular}

\end{table}

\subsection{Results: Murmur detection using PCG 2022 dataset}

{\bf Remark:}
Our experimental results indicate that the wavelet scattering transform (WST) is the most effective feature extraction method that helps us differentiate between the three classes for the CirCor Digiscope 2022 dataset (the same holds for the PCG 2016 dataset). Therefore, we report the results due to WST only, unless otherwise specified. 

Next, recall that experiments E1, E2, E3 do murmur detection using PCG 2022 dataset as is, with unknown class removed, with re-labeling of segments, respectively. 

{\it Performance of 1D-CNN:}
We first report the results obtained by the best-performing 1D-CNN model. For experiment E1, we obtained an accuracy of 74.39\%, weighted accuracy of 78.06\%, F1 score of 62.3\%, and area under the receiver operating characteristic curve (AUROC) of 82.11\% (see Table \ref{table:sota}). 
For experiment E2, we obtained an improved accuracy of 90.09\%, elevated F1-score of 90.09\%, and higher AUROC of 95.32\%. Finally, for experiment E3, the weighted accuracy improved to 83.69\% and AUROC rose to 91.82\%. 

Table \ref{table:E1-E3} shows the detailed confusion matrices for the experiments E1-E3. We observe that experiment E3 leads to a more balanced confusion matrix, compared to experiment E1. That is, one can see that only 18.64\% of the unknown class samples were correctly classified in experiment E1. But in experiment E3 (after re-labeling of noise-only segments), the number of true positives increases to 78.03\%. This phenomenon could also be verified by comparing the F1-scores for experiments E1 and E3 in Table \ref{table:sota}. 

{\it Selection of optimal segment size:}
We repeated our experiment E1 (where we utilize the original PCG 2022 dataset as is) with different segment sizes (i.e., 1 sec, 3 sec, 4 sec, and 5s), and assessed the performance of our 1D-CNN model with the aim to find an optimal segment size (see Table \ref{table:window_impact}). We utilize F1-score as the core performance metric to select the optimal segment size (this is because the accuracy is not a reliable performance metric for the imbalanced datasets). Furthermore, we observed that overlapping segments cause extreme over-fitting; therefore, we ended up selecting non-overlapping segments of duration 4 seconds. 

%so compared with 4s before cleaning as well. 

\begin{table}[h]
    \caption{Impact of different segment sizes on performance of our 1D-CNN model (for experiment E1). }
    \centering
    \begin{tabular}{  |>{\centering\arraybackslash}p{1.5cm} | >{\centering\arraybackslash}p{1.25cm} | >{\centering\arraybackslash}p{1.25cm} | >{\centering\arraybackslash}p{1.25cm} | >{\centering\arraybackslash}p{1.25cm}|  }
        \hline
        Window size  & Accuracy & Precision & Recall & F1-score  \\
        \hline
        1 sec & 72.15\% & 55.40\% & 54.91\% & 54\%    \\
        % \hline
        3 sec & 86   \% & 55.60\% & 55.30\% & 55.30\% \\
        % \hline
        4 sec & 74.39\% & 62.69\% & 62\% & 62.3\% \\
        % \hline
        5 sec & 71.20\% & 62.32\% & 60.91\% & 61.6\% \\
        \hline

    \end{tabular}
    \label{table:window_impact}
\end{table}

\begin{table*}[h]
    \caption{ Heart murmur detection using PCG 2022 dataset: first box does performance comparison of our work with the state-of-the-art. The second box presents the performance of our 1D-CNN model for experiment E2. The third box shows the performance of our 1D-CNN, LSTM-RNN, and CRNN models for experiment E3. }
    \centering
    \begin{tabular}{ | p{4.7cm} | p{1.5cm} | p{2.5cm} | p{1.5cm} | p{1.5cm} | p{1.5cm} | p{1.5cm} | }
        \hline
        Works    & Accuracy    & Weighted Accuracy & Precision & Recall & F1-Score & AUROC  \\
        \hline
        \cite{rank1} & 80.1\%     & 78\%            & -----   & ----- & 61.9\% & 88.4\%  \\
        \hline
        \cite{rank2} & 76.3\%     & 77.6\%          & -----   & ----- & 62.3\% & 75.7\%  \\
        \hline
        \cite{rank3} & 82.2\%     & 77.6\%          & -----   & ----- & 64.7\% & 77.1\%  \\
        \hline
        Our work: 1D-CNN (experiment E1) & 74.39\%    & 78.06\%       & 62.69\% & 62 \% & 62.3\% & 82.11\%  \\
        \hline
        {Our work: 1D-CNN with voting (experiment E1)} & -----    & -----       & 69.12\% & 65.24\% & 65.5\% & -----  \\
        \hline \hline
        Our work: 1D-CNN (experiment E2) & 90.09\%         & -----                   & 90.12\% & 90.14\% & 90.09\% & 95.32\%  \\
        \hline
        {Our work: 1D-CNN with voting (experiment E2)} & -----         & -----                   & 88.87\% & 91.12\% & 90.0\% & -----  \\
        \hline \hline 
        Our work: 1D-CNN (experiment E3)  & 78.7\%    & 83.69\%       & 79.34\% & 78.69\% & 78.67\% & 91.82\%  \\
        \hline
        {Our work: 1D-CNN with voting (experiment E3)}  & -----    & -----       & 76.41\% & 77.82\% & 76.54\% & -----  \\
        \hline
        Our work: LSTM-RNN (experiment E3)  & 72.28\%    &78.42\%       & 73.62\% & 72.54\% & 72.60\% & 88\%  \\
        \hline
        Our work: CRNN (experiment E3)   & 74.39\%    & 81.98\%       & 75.30\% & 74.10\% & 74.31\% & 88.93\%  \\
        \hline
        Our work: 1D-CNN  (experiment E3) - MFCC features   & 77.41\%    & 81.51\%       & 78.07\% & 77.34\% & 77.30\% & 91.10\%  \\
        \hline
        Our work: 1D-CNN  (experiment E3) - STFT features   & 72.33\%    & 76.95\%       & 74.22\% & 72.50\% & 72.28\% & 86.72\%  \\
        \hline

    \end{tabular}
    \label{table:sota}
\end{table*}

{\it Performance comparison of our 1D-CNN model with LSTM-RNN and CRNN models:}
Table \ref{table:sota} lists the performance obtained by the other two NN models (i.e., LSTM-RNN and CRNN) that we implemented and tested, for experiment E3 (which does murmur detection on the re-labeled dataset). We note that our 1D-CNN model outperforms both LSTM-RNN and CRNN models in terms of accuracy, weighted accuracy, precision, recall, F1-score and AUROC. 

{\it Performance comparison with the state-of-the-art:}
Table \ref{table:sota} provides a thorough comparison of the performance achieved by our NN models (i.e., 1D-CNN, LSTM-RNN, and CRNN) with the performance achieved by the top three works \cite{rank1,rank2,rank3}. Table \ref{table:sota} demonstrates that our 1D-CNN model outperforms the works \cite{rank1,rank2,rank3} in terms of accuracy, weighted accuracy, F1-score and AUROC, for experiment E3 (which does murmur detection on the cleaned dataset). As for experiment E1 (which does murmur detection on the original dataset), we see that our model performs very close to \cite{rank1} in terms of weighted accuracy, and to \cite{rank2} in terms of F1-score.

{\it Performance of the voting-based approach:}
We also implemented the voting-based approach for experiments E1-E3. Under this method, we simply group all samples in the test split that belong to the same heart auscultation location of a given patient. We then inspect the classification result of each sample, and finally choose the label that has the maximum number of votes. Note that grouping all segments that belong to the same chest location in the same person means a change in data distribution. That is, there might be longer recordings for "present" class and shorter recordings for "absent" class, but then after the grouping, the number of "present" samples will go down, since most of the samples got grouped together. 
For this reason, we use F1-score as the performance metric.
For experiment E1, we obtain an F1 score of 65.5\%, precision of 69.12\%, and recall of 65.24\%. Experiment E2 achieves an F1-score of  90\%, precision of 88.87\%, and recall of 91.12\%. 
Finally, for experiment E3, we achieve an F1-score of 76.54\%, precision of 76.41\%, and recall of 77.82\%. Thus, to sum things up, voting-based approach leads to some performance boost for experiment E1, no change for experiment E2, and slight performance degradation for experiment E3. 
The detailed confusion matrices of our 1D-CNN model with voting-based approach for experiments E1-E3 are shown below in Table \ref{table:voting}. 

{\it Performance of MFCC and STFT feature extraction methods:}
{For the sake of completeness, Table \ref{table:sota} also summarizes the results obtained by the vanilla 1D-CNN when it utilizes the feature vectors from other feature extraction methods (i.e. MFCC and STFT), for experiment E3. We observe that the MFCC-based feature extraction performs close to the Wavelet Scattering method, but the performance of STFT based feature extraction method is below par.}

%\begin{figure}[htp]
 %   \centering
  %  \subfloat[]{\includegraphics[width = 1.5in]{Confusion-Matrices/cf_murmur_3classes_reannotated.png}}\\
   % \caption{ confusion matrix for CirCor dataset after dataset cleaning }
    %\label{fig:cf-2022-reannotated}
%\end{figure}

%Table \ref{table:Clinical-outcome-results} provides the clinical outcome performance comparison.

%\begin{table}[h]
 %   \caption{Results of clinical outcome using PCG 2022 dataset after manual cleaning, compared to the top performing submissions to the PhysioNet-2022 challenge}
  %  \centering
   % \begin{tabular}{ | p{2.5cm} p{2.5cm} | }
    %    \hline
     %   Works    & Clinical Cost Result  \\
      %  \hline
       % Rank \#1 contestant & 11144\\
        %\hline
        %Rank \#2 contestant & 11403\\
        %\hline
        %Rank \#3 contestant & 11735\\
        %\hline
        %Our work: 1D-CNN    & 17811\\
        %\hline

    %\end{tabular}
    %\label{table:Clinical-outcome-results}
%\end{table}

\subsection{Results: Abnormal PCG detection using PCG 2016 dataset}

\begin{figure}[htp]
    \centering
    \subfloat[]{\includegraphics[width = 1.5in]{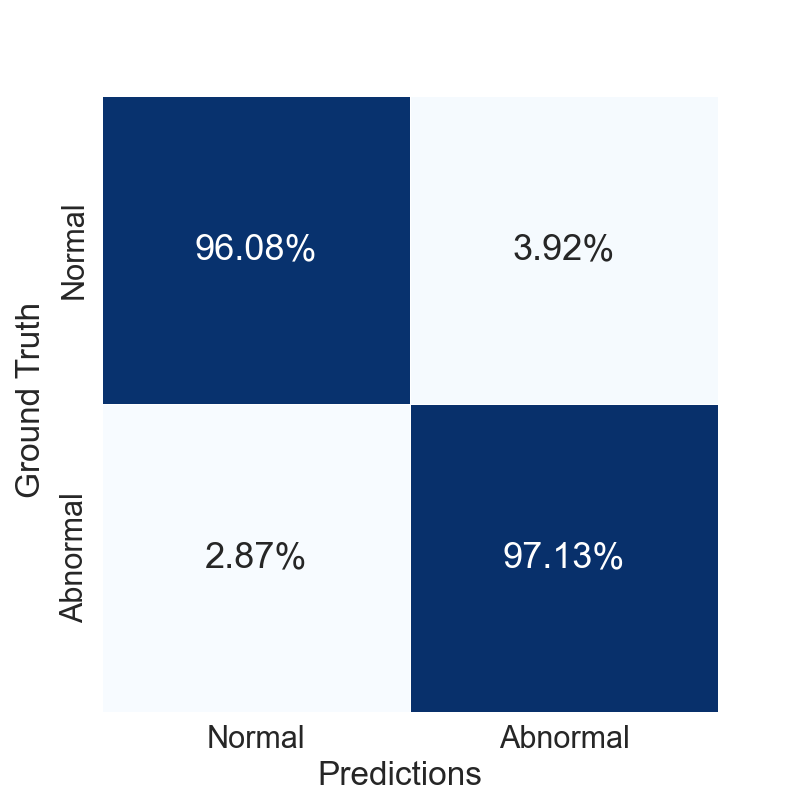}}\\
    \caption{ Confusion matrix for experiment E4: abnormal PCG detection (using PCG 2016 dataset). }
    \label{fig:cf-2016}
\end{figure}

As mentioned in the previous section, we re-train our 1D-CNN model from scratch on PCG 2016 dataset in order to differentiate between a normal PCG signal and an abnormal PCG signal. The hyper-parameters of the 1D-CNN that was fine-tuned for PCG 2016 dataset could be found in Table \ref{table:HyperParameters}. For the abnormal PCG detection problem, the model achieved an accuracy of 96.51\%, precision of 96.26\%, recall of 96.60\%, F1 score of 96.42\%, and AUROC of 98.17\%. Thus, our custum 1D-CNN model outperforms the top performing method, i.e., state-of-the-art method reported in \cite{2016challengewinner} (see Table \ref{table:2016-results}). Additionally, Fig. \ref{fig:cf-2016} provides confusion matrix that outlines the true positives, true negatives, false positives, and false negatives. 

% Table Below was too big to fit in a column, so I rewrote it below in two columns instead
% \begin{table}[h]
%     \caption{Results of murmur detection using 2022-CirCor dataset after manual cleaning, compared to the top performing submissions to the PhysioNet-2022 challenge (evaluated on test set)}
%     \centering
%     \begin{tabular}{ | p{2cm} p{0.75cm} p{1cm} p{0.75cm} p{0.75cm} p{0.75cm} p{0.75cm} | }
%         \hline
%         Team    & Accuracy    & Weighted Accuracy & Precision & Recall & F1-Score & AUROC  \\
%         \hline
%         Rank \#1 & 80.1\%     & 78\%            & -----   & ----- & 61.9\% & 88.4\%  \\
%         \hline
%         Rank \#2 & 76.3\%     & 77.6\%          & -----   & ----- & 62.3\% & 75.7\%  \\
%         \hline
%         Rank \#3 & 82.2\%     & 77.6\%          & -----   & ----- & 64.7\% & 77.1\%  \\
%         \hline
%         CNN1D (Before dataset cleaning) & 82.28\%    & 83.81\%       & 65.67\% & 65.92\% & 65.79\% & 90.79\%  \\
%         \hline
%         CNN1D (After dataset cleaning)  & 82.86\%    & 86.30\%       & 81.75\% & 82.19\% & 81.87\% & 93.45\%  \\
%         \hline
%         LSTM (After dataset cleaning)  & 77.18\%    & 82.28\%       & 76.83\% & 76.20\% & 75.92\% & 90.72\%  \\
%         \hline
%         CRNN. (After dataset cleaning)  & 73.99\%    & 79.88\%       & 74.02\% & 74.38\% & 73.64\% & 89.25\%  \\
%         \hline

%     \end{tabular}
%     \label{table:murmur-reannotated-results}
% \end{table}

\begin{table}[h]
    \caption{Experiment E4 (Abnormal PCG signal detection): performance of our 1D-CNN model on PCG 2016 dataset. }
    \centering
    \begin{tabular}{ | p{1.2cm} p{1cm} p{1cm} p{1cm} p{1.5cm} p{1cm} | }
        \hline
        Work & Accuracy & Precision & Recall & F1-Score & AUROC  \\
        \hline
        {Work \cite{2016challengewinner} } & 92.59\%  & 92.53\%   & 92.59\% & 92.54\% & 98\%  \\
        \hline
        Our work & 96.51\%  & 96.26\%   & 96.60\% & 96.42\% & 99.10\%  \\
        \hline
    \end{tabular}
    \label{table:2016-results}
\end{table}

% \begin{figure}[htp]
%     \centering
%     \subfloat[]{\includegraphics[width = 1.2in]{Confusion-Matrices/Rev1/voting system/Exp1.png}} \hspace{0.1in} 
%     \subfloat[]{\includegraphics[width = 1.2in]{Confusion-Matrices/Rev1/voting system/Exp2.png}} \hspace{0.1in}
%     \subfloat[]{\includegraphics[width = 1.2in]{Confusion-Matrices/Rev1/voting system/Exp3.png}}\\
%     \caption{ Confusion matrices when applying voting system. \\
%     (a) Exp1\\
%     (b) Exp2 \\
%     (c) Exp3
%     }
%     \label{fig:Exp1-2-3 voting system confmatrix}
% \end{figure}

\begin{table}[h!]
\centering
\caption{Confusion matrices of our 1D-CNN model due to voting-based approach (for experiments 1, 2, and 3).}
\label{table:voting}
\begin{tabular}{|c| c| c| c|c|} 
 \hline
 Experiment &
	Actual /Predicted&	Absent &	Unknown	& Present
\\

\hline
     &	Absent  &	\cellcolor{blue!25}91.61\% &                      2.80\% &	                   5.59\%  \\
E1 &	 Unknown &	                    85\%    &	\cellcolor{blue!25}15\%   &	                    0\%     \\
     &	Present &	                   8.70\%  &	                  2.17\% &	\cellcolor{blue!25}89.13\% \\
 
\hline
     &	Absent  &	\cellcolor{blue!25}94.64\% &                           - &	                   5.36\%  \\
E2 &	 Unknown &	                          - &	                        - &	                          - \\
     &	Present &	                   12.07\% &                           - &	\cellcolor{blue!25}87.93\% \\
     
\hline
     &	Absent  &	\cellcolor{blue!25}79.31\% &                      17.82\% &	                   2.87\% \\
E3 &	 Unknown &	                    28.07\% &	\cellcolor{blue!25}71.93\% &	                    0\% \\
     &	Present &	                   6.67\% &	                      11.11\% &	\cellcolor{blue!25}82.22\% \\

 \hline
\end{tabular}
\end{table}

\section{Conclusion}

This work utilized two public datasets, i.e., CirCor Digiscope 2022 dataset and PCG 2016 datasets (from Physionet online database) in order to train three deep learning classifiers, i.e., 1D-CNN, LSTM-RNN, and C-RNN to do heart murmur detection as well as abnormal PCG detection. 
We observed that our 1D-CNN outperformed the other two NNs, with an accuracy of 78.7\%, weighted accuracy of 83.69\%, and F1-score of 78.67\% (for experiment E3 which does murmur detection after re-labeling of noise-only segments). The outcomes of this work could assist doctors, and could lead to potential clinical workflow improvements ultimately leading to autonomous diagnosis.

We note that the CirCor Digiscope 2022 dataset also contains other valuable information about the murmurs, e.g., murmur grading, timing, quality, pitch, shape etc. Thus, design of novel machine/deep learning algorithms which could do automatic and accurate murmur grading analysis with little data is an open problem of great interest. Furthermore, study of generative methods which could reliably generate synthetic PCG data in order to help train the data hungry deep learning methods is another interesting but challenging open problem (due to non-stationary nature of PCG signals with murmurs).

\section{Appendix A: computation of MFCC features}

The process of computation of MFCC consists of a number of steps, as follows.
The first step is to apply a pre-emphasis filter to the signal \( x[n] \):
\begin{equation}
    y[n] = x[n] - \alpha x[n-1],
\end{equation}
where \( \alpha \) is typically between 0.95 and 0.97. Next, the signal is divided into overlapping frames. If the frame size is \( N \) and the hop size is \( M \), then each frame can be represented as:
\begin{equation}
    x_m[n] = x[n + mM], \quad m = 0, 1, 2, \ldots
\end{equation}
Each frame is then windowed using a window function \( w[n] \), such as the Hamming window:
\begin{equation}
    x_w[n] = x[n] \cdot w[n], \quad w[n] = 0.54 - 0.46 \cos\left(\frac{2\pi n}{N-1}\right)
\end{equation}
We then Compute the fast Fourier transform (FFT) of each windowed frame to obtain the magnitude spectrum:
\begin{equation}
    X[k] = \sum_{n=0}^{N-1} x_w[n] e^{-j2\pi k n / N}, \quad k = 0, 1, \ldots, N-1
\end{equation}
Next, we apply the mel filter bank to the magnitude spectrum to get the filter bank energies:
\begin{equation}
    E_m = \sum_{k=0}^{N-1} |X[k]|^2 H_m[k], \quad m = 1, 2, \ldots, M
\end{equation}
where \( H_m[k] \) is the mel filter bank.
We then proceed to take the logarithm of the filter bank energies:
\begin{equation}
    F_m = \log(E_m), \quad m = 1, 2, \ldots, M
\end{equation}
Finally, we apply the discrete Cosine transform to the logarithm of the filter bank energies to obtain the MFCCs:
\begin{equation}
    C_n = \sum_{m=0}^{M-1} F_m \cos\left[\frac{\pi n}{M}\left(m + \frac{1}{2}\right)\right], \quad n = 0, 1, \ldots, L-1
\end{equation}
where \( L \) is the number of desired MFCC features.

% or


%\appendices

%\section{}

%\section*{Acknowledgment}


\ifCLASSOPTIONcaptionsoff
  \newpage
\fi



%\bibliographystyle{IEEEtran}

\bibliographystyle{IEEEtran}
\bibliography{refs}

\end{document}